

\def\frac#1#2{{#1\over#2}}

\def\sqr#1#2{{\vcenter{\vbox{\hrule height.#2pt
   \hbox{\vrule width.#2pt height#1pt \kern#1pt
   \vrule width.#2pt}
   \hrule height.#2pt}}}}
\def\square{{\mathchoice\sqr65\sqr65\sqr54\sqr{3.5}4}\,}

\magnification=\magstep1
\openup1\jot
\vsize=9truein
\font\titlefont=cmcsc10 scaled \magstep2
\hskip290pt{DAMTP-R93-12}
\smallskip\hskip300pt{CALT-68-1861}\bigskip
\centerline {\titlefont Quantum Coherence in Two Dimensions }
\vskip 1truein
\centerline {S.~W.~Hawking}\centerline{ \&} \centerline{ J.~D.~Hayward}
\vskip .5truein
\centerline {Department of Applied Mathematics and Theoretical Physics}
\centerline {University of Cambridge}
\centerline {Silver Street}
\centerline {Cambridge CB3 9EW}
\centerline {UK}
\vskip .25truein\centerline{ \&}\centerline{ California Institute of
Technology}
\centerline{ Pasadena}\centerline{ California 91125 USA}
\centerline {\it March 1993  }
\vskip 0.5truein
\centerline {\bf Abstract}

\bigskip

The formation and evaporation of two dimensional black holes are discussed.
 It is shown that if the radiation in minimal scalars has positive energy,
there must be a global event horizon or a naked singularity. The former would
imply loss of quantum coherence while the latter would lead to an even worse
 breakdown of predictability. CPT invariance would suggest that there ought
 to be past horizons as well. A way in which this could happen with wormholes
 is described.
\footnote{}{ S.W.Hawking@amtp.cam.ac.uk, J.D.Hayward@amtp.cam.ac.uk}
 \vfill \eject
 \beginsection 1. Introduction

 The discovery that black holes emit radiation [1] suggests that they  will
evaporate and
 eventually disappear. In this process it seems that information  and quantum
coherence will
 be lost and the evolution from initial to final situation will be described
not by an S matrix
 acting on states but by a super scattering operator \$ acting on density
matrices [2]. This proposal
 of a non unitary evolution evoked howls of protest when it was first put
forward and three possible
 ways of maintaining the purity of quantum states were put forward:
 \item 1 The apparent horizon  eventually disappears and allows the information
that
 went into the hole to return.
 \item 2 The back reaction to the emission of radiation introduces subtle
 correlations between the different modes. These allow the information to
 come out continuously as the black hole evaporates.
 \item 3 The black hole does not evaporate completely but leaves some small
 remnant that still contains the information.

 The first possibility, that the information comes out at the end of the
evaporation,
 has the difficulty that energy is required to carry the information remaining
in the
 black hole. However, there is very little rest mass energy left in the final
stages
 of the evaporation.
The information can therefore be released only very slowly, and one has a long
lived
 remnant, like in possibility three.

The second possibility, that the information comes out continuously during the
evaporation, has problems with causality. The particles falling into the hole
would
 carry their information far beyond the horizon before the curvature would
become
 strong enough for quantum gravitational effects to be important. Yet the
information
 is supposed to appear outside the apparent horizon. If one could send
information
 faster than light like that, one could also send information back in time,
with
 all the difficulties that would cause.

The third possibility, black hole remnants, has problems with CPT if black
holes
could form but never disappear completely.
Consider a certain amount of energy placed in a box with reflecting walls[3].
The energy can
be distributed in a large number of microscopic configurations, but one of two
situations
will correspond to the great majority: either just thermal radiation, or
thermal radiation
in equilibrium with a black hole at the same temperature. Which possibility has
more phase
space depends on the energy and the volume of the box.

Suppose the energy is sufficiently low
and the volume sufficiently large that just thermal radiation, with no black
hole, corresponded
to more states. Then for most of the time there would be no black hole in the
box. However, occasionally a
 black hole would form by thermal fluctuations, and then evaporate
again. By CPT one would expect this process to be time symmetric. That is, if
you took a film,
it would look the same running forwards and backwards. But this is impossible
if black holes
can form from nothing but leave remnants when they evaporate. One can not even
restore CPT, and
get a sensible picture, by supposing there's a separate species of white holes
that would have
existed from the beginning of time. The number of white holes would always be
going down, and
the number of black hole remnants would be going up, so one could never have a
statistical
equilibrium in the box.
 We shall have more to say about CPT later.
 It is difficult to see how information and quantum coherence could be
preserved in
 gravitational collapse. However, because General Relativity
 is non renormalizable, it is not clear what will happen in the final stages of
black
hole evaporation. Thus the question of whether quantum coherence is lost is
still open.
 For this reason there has recently been interest in two dimensional theories
of quantum
 gravity which show an analogue of black hole radiation and which have the
great advantage
 of being renormalizable.

 The first two dimensional theory that could describe the formation and
evaporation of
 black holes was put forward by Callan, Giddings, Harvey and Strominger (CGHS)
[4].
 It contained a metric $g$ and a dilaton $\phi $ coupled  to $N$ minimal scalar
fields
$f \sb i$. In the classical theory a black hole can be created by sending a
wave of one
 of the scalar fields. Quantum theory on this classical black hole background
then predicts
 the black hole will radiate at a steady rate indefinitely. CGHS hoped that the
inclusion
of the back reaction would cause  the field configuration that initially
resembled a black
 hole to disappear without a singularity or a global event horizon. Thus they
hoped there
would be no loss of information and hence no loss of quantum coherence.

 However, the most straightforward inclusion of the back reaction in the semi
classical
 equations did not realize this hope. There was necessarily a singularity where
the dilaton
 had a certain critical value [5][6]. This singularity could either become
naked, that is,
 visible from future null infinity at late retarded times [7][8][9] or it could
be a
thunder-bolt that cut off future null infinity at a finite retarded time
[10][11].
In either case part of the information about the initial quantum state would be
lost on
the singularity, which would be space like for at least part of its length, so
one might
 expect loss of quantum coherence. The back reaction used in these calculations
is based
on the obvious and unambiguous measure for the path integral over the minimal
scalars and
the ghosts but it is not so clear what measure to use for the dilaton and the
conformal
factor. In the large $N$ limit this ambiguity in the measure shouldn't matter
but the main
 hope of would-be defenders of quantum purity was that the large quantum
fluctuations when
the dilaton was near its critical value would cause the large $N$ approximation
to break
down and that higher order quantum corrections might prevent the occurence of
singularities
 and preserve quantum coherence. However, in this paper it will be shown that
if the
 emission in scalar has positive energy, then there must be either naked
singularities
 or event horizons or both. This argument depends only on the known measure for
the
minimal scalars, and is independent of any corrections to the equations of
motion that
may arise from the measure on the dilaton and conformal factor or from higher
order
quantum effects.

 \beginsection 2. The conservation equations

 The argument is based on the fact that the conservation equations and the
trace anomaly
of the scalar fields determine their energy momentum tensor up to constants of
integration
 which can be fixed by boundary conditions. In the conformal gauge in which the
metric is
 $$ds^2=-e^{2 \rho} dx_+ dx_-\eqno(1)$$ the energy momentum tensor of each of
the minimal
 scalars is $$T_{\pm\pm}=-{1\over{12}}\left( \left({\partial \rho \over
\partial x_{\pm}}\right)^2-{\partial ^2\rho \over \partial
{x_{\pm}}^2}+t_{\pm}(x_{\pm})\right) \eqno(2)$$
 $$T_{+-}=-{1\over{12}}\partial_{+}\partial_{-}\rho \eqno(3)$$ where
$t_{\pm}(x_{\pm})$ are
 constants of integration.

 Consider a situation in which the spacetime is flat, so that the conformal
factor is of the form
$\rho = \log F(x_-)+\log G(x_+) $ and the energy momentum is zero before some
null geodesic $\gamma $. This would be the case if the initial state was the
linear dilaton
 solution. On the null geodesic $\gamma$ one can change the coordinate $x_-$ to
$\int^{x_-} F^2 dx_-^{\prime}$ so that $\rho =0$ on $\gamma $. The range of
$x_-$ will be $(-\infty ,\infty )$. From the assumption that the energy
 momentum tensor is zero initially, it then follows that $t_-(x_-)=0$ for all
$x_-$.

 Suppose now that a wave with positive energy is sent in from the asymptotic
region of weak
 coupling at an advanced time $x_+$ later than $\gamma $ and creates some black
hole like object
 which radiates energy in the $N$ minimal scalar fields. By equation $(2)$, the
outgoing energy flux
in the minimal scalar fields will be $${\cal E}={N\over 12}\left( {\partial
^2\rho \over \partial
 {x_-}^2}- \left({\partial \rho \over \partial x_-}
\right)^2 \right)  \eqno(4)$$ Let $\lambda $ be an ingoing null geodesic at
late advanced time.
 If the outgoing energy flux crossing $\lambda $ is non negative, $${\partial
^2\rho \over \partial
 {x_-}^2}\ge \left({\partial \rho \over \partial x_-}\right)^2\eqno(5)$$

 To integrate $(5)$ along $\lambda $, one needs to know the initial value of
 $\partial \rho /\partial x_-$. Let $\mu $ be an outgoing null geodesic from a
point $p$ on
 $\gamma $ to a point $q$ on $\lambda $. We shall choose $\mu $ to lie in the
asymptotic region,
 that is, at early retarded times. One can choose the $x_+$ coordinate along
$\mu $ so that $\rho =0$
 on $\mu $. This fixes the coordinates up to a Poincare transformation. With
this choice of coordinates,
 $${\partial \rho \over \partial x_-}\big|_q ={1\over 8}\int_p^q R\,
dx_+\eqno(6)$$ One would expect the
curvature $R$ on $\mu $ to be positive and exponentially decreasing if the
Bondi mass measured at infinity,
$$ M\propto e^{-2\phi}R\vert_{x_-\to -\infty} \eqno(7)$$
on $\mu $ is positive. Thus, if one takes the null geodesic $\mu $ to be
sufficiently far out in the
 asymptotic region, the integral $(6)$ will be positive.

 Suppose now that the outgoing energy flux $T_{--}$ is strictly positive on
some interval of $\lambda $
 around a point $r$ to the future of $q$. Then it follows from $(5)$ and $(6)$
that to the future of $r$
 on $\mu $ $$ \rho \ge \log (c-b)-\log (c-x_-)\eqno(8)$$ where $b$ is the value
of $x_-$ at $r$ and $c$
 is some finite quantity greater than $b$. From $(8)$ it follows that $\rho $
will diverge at some point
 $s$ on $\mu $ where $x_-=a \le c$. The point $s$ may or not be singular in the
sense of the curvature
 $R$ being unbounded but it will necessarily be at an infinite affine parameter
distance along $\lambda  $.
 It will however be at a finite retarded time $x_-$ (Fig 1). This means that
the original hope of CGHS, that the
 black hole would evaporate without global horizons or singularities, can not
be realized in any two
 dimensional quantum theory in which the energy emission is positive.

 Let $\bar {\lambda }  $ be the portion of $\lambda  $ up to $s$. Then
$J^-(\bar {\lambda })$, the past of
 $\bar {\lambda } $, will not include the whole of the null geodesic, $\gamma
$, in the initially flat region.
 It is this kind behavior that gives rise to thermal radiation. Let $\bar
h(x_-)$ be a wave packet that is zero
 for $x_->a $ and is purely positive frequency with respect to the affine
parameter on the late time null
 geodesic $\bar {\lambda }$. Then $\bar h(x_-)$ is not purely positive
frequency with respect to the affine
 parameter on $\gamma $ (which is proportional to $x_-$) because it is zero in
a semi infinite range.
Instead, there will be some wave packet $\hat h(x_-)$ which is zero for $x_-<a
$ and which is such
 that $\bar h+\hat h $ is purely positive frequency on $\gamma $. This will
mean that the initial vacuum state
in each of the minimal scalar fields $f_i$ will appear to contain pairs of
particles, one in the $\bar h $ mode,
 and the other in the $\hat h$ mode. The $\bar h $ mode will appear to contain
a particle on the null geodesic
 $\bar {\lambda }$. But the $\hat h $ will not cross $\bar {\lambda }$, so an
observer in the asymptotic region

 will not see this particle. This would mean that the quantum state would
appear to be a mixed state, described
 by a density matrix obtained by tracing out over the modes for $x_->a $. Thus
there will be loss of quantum
 coherence.

 In the above, we have implicitly assumed that every outgoing null geodesic
that intersects $\bar {\lambda }$,
 also intersects $\gamma $. This allows us to deduce that the constant of
integration $t_-(x_-)=0$ on each
 outgoing null geodesic. However, if there was a singularity that was naked in
the sense that it was visible
 from $\bar {\lambda }$, it wouldn't follow that on $\bar {\lambda }$
$${\partial ^2\rho \over \partial x_-^2}\ge \left ({\partial \rho \over
\partial x_-}\right)^2$$
 Thus the requirement that the radiated energy is positive implies either that
there is an horizon and
 associated loss of quantum coherence, or there is a naked singularity. In our
opinion, this would be much
 worse.

 The discussion so far has been in terms of a semi classical metric. However it
should also apply to each
 individual metric in a path integral over all metrics and dilaton field
because our conclusions depend only
 on the asymptotic form of the metric in the far future and past. Thus we would
expect loss of quantum
coherence, or naked singularities, or both, in the full quantum theory.

 \beginsection 3. Conformal Treatment of Infinity

 In the previous discussion, the null geodesic $\gamma $ was at early advanced
time, the null geodesic
 $\lambda  $ was at late advanced time, and the null geodesic $\mu  $ that
connected them was at early
retarded time. To make the arguments about the positive mass and energy of the
emitted radiation, one wants
 to take the limit that these three null geodesics are at infinitely early or
late advanced or retarded
times. A precise and elegant way of doing this is to use the concept of
conformal infinity that was introduced
 by Penrose in the four dimensional case. One takes the spacetime manifold and
metric $M,g\sb {\mu \nu}$ to
 be conformal to a manifold with boundary and conformal metric $\tilde M,\tilde
g\sb {\mu \nu}$ where $$g\sb {\mu \nu}=\Omega ^{-2}\tilde g\sb {\mu \nu}$$
$$\Omega =0~~~~{\rm on~}\partial \tilde M$$

 The curvature scalars of the two metrics are related by $$ R=\Omega^{2}\tilde
R+2\Omega
\tilde {\square}\Omega-2(\tilde {\nabla}\Omega)^2 \eqno(9) $$ where the
covariant derivatives are with
 respect to the conformal metric $\tilde g\sb {\mu \nu}$. The physical
curvature $R$ will go rapidly to zero
 in the weak coupling region. It then follows from (9) that the boundary
$\partial \tilde M$ will be null
 where $\nabla \sb {\mu}\Omega \ne 0$. The boundary in the weak coupling region
can be divided into future
 and past weak null infinities ${\cal I}^{\pm}_w$. They will be joined by the
point $I ^0$ representing
spatial infinity. The conformal factor $\Omega $ will not be smooth at $I ^0$.
One can not say anything in
 general about the part of the $\partial \tilde M $ that lies in the strong
coupling region because one does
 not know how $R$ will behave there. However, in the case that spacetime is
flat before some ingoing null
geodesic $\gamma $, one will have a past strong null infinity ${\cal I} ^-_s$,
but one can not assume that
there is necessarily a future strong null infinity.

 One can take the conformal metric $\tilde g\sb {\mu \nu}$ to be flat. Then one
can take $\tilde M $ to be
 the region in two dimensional Minkowski space bounded by three null geodesics
${\cal I} ^-_s$, ${\cal I} ^-_w$
 and ${\cal I} ^+_w$
 (Fig 2). One does not know the form of the boundary on the fourth side, but
this does not matter for the problem
 under consideration.

 The quantity $\tilde \rho =-\log \Omega $ will differ by a solution of the
wave equation from the $\rho $
 used in the previous section since it will obey different boundary conditions:
$\tilde \rho =\infty$ on
$\partial \tilde M$ while $\rho =0$ on $\gamma $ and $\lambda $. In order to
identify the coordinate
independent part of $\rho $ and $\tilde \rho $ we shall introduce a field $Z$
with the coupling
$$\square Z = -\nu R\eqno(10) $$ $$\tilde {\square} Z = -\nu \Omega
^{-2}R\eqno(11) $$ We shall assume that the
 physical curvature goes to zero fast enough that $\Omega ^{-2}R$ is bounded on
${\cal I} ^+_s$ and ${\cal I} ^+_w$.
 One can then solve the wave equation $(3)$ on the conformal spacetime $(\tilde
M, \tilde g \sb {\mu \nu})$ with the
boundary conditions that $Z=0$ on ${\cal I} ^-_s$ and ${\cal I} ^-_w$. The
field $Z$ on $M$ obtained in this way
 will correspond to $2\nu \rho $ where $\rho $ is the conformal factor in the
previous  section in the limit that
 the null geodesic $ \mu $ is taken to infinity.

 The energy momentum tensor of the $Z$ $$T_{\mu\nu}={1\over 2}(\nabla_{\mu}
Z\nabla_{\nu} Z-{1\over 2}g_{\mu\nu}
(\nabla Z)^2)
+ \nu(\nabla_{\mu}\nabla_{\nu} Z-g_{\mu\nu}\square Z)\eqno(12) $$ will
correspond to the energy momentum of the
 radiation in the
 $N$ minimal scalar fields if $\nu ^2=N/24$. Thus the energy out flow across
${\cal I} ^+_w$ is
$${\cal E}=T_{\mu \nu}n^{\mu}n^{\nu}= {1\over 2}\left(\nabla
_{\mu}Zn^{\mu}\right)^2+\nu \nabla _{\mu}\nabla _{\nu}
Zn^{\mu}n^{\nu}\eqno(13)$$ $$={1\over 2}\left({dZ\over dt}\right)^2+\nu
\left({d^2Z\over dt^2} -q{dZ\over dt}\right)
 \eqno(14)$$ where $n^{\mu}=dx^{\mu}/dt$ is the tangent vector to ${\cal I}
^+_w$, $t$ is a parameter
 along ${\cal I} ^+_w$ and $n^{\nu}\nabla_{\nu}n^{\mu}=q\, n^{\mu}$.

Define a metric $\hat g _{\mu \nu}=\exp (-Z\nu ^{-1})g_{\mu \nu} $. This metric
is flat and corresponds to the
 flat background metric in section 2 in the limit that the null geodesic $\mu $
is taken to infinitely early
retarded times. Let $t$ be an affine parameter with respect to the metric $\hat
g $ on ingoing null geodesics.
 Because $\hat g$ is flat, one can choose $t$ to be constant on each out going
null geodesic.

Near ${\cal I} _s^-$, $Z=0$ and the range of $t $ will be $(-\infty,\infty)$.
At later advanced times,
 $Z\ne 0$ and $$q=\nu ^{-1}{dZ\over dt} \eqno(15)$$ Thus the energy flux across
${\cal I}_w^+$
is $${\cal E}= -{1\over 2}\left({dZ\over dt}\right)^2+ \nu {d^2Z\over dt^2}
\eqno(16)$$ If one
 replaces $Z$ with $2\nu \rho $, (16) becomes the same as (4). If the mass
measured on ${\cal I}_w^-$
 is positive, $R\ge0$ near ${\cal I}_w^-$. If $\nu >0$, this implies $Z\ge 0$
and ${dZ\over dt}\ge 0$
 near ${\cal I}_w^-$.

The argument is now similar to that in section 2. If $\cal E$ is non negative
on ${\cal I}_w^+$ and is strictly
 positive on some interval, then by (16), $Z$ will diverge at a point $s$ on
${\cal I}_w^+$ at a finite value
 of the parameter $t$. But the range of $t$ on ${\cal I}_s^-$ is infinite. Thus
there will be a part of
 ${\cal I}_s^-$ which is not in the past of $s$ which is the future end point
of ${\cal I}_w^+$ because
it is at infinite distance in the natural affine parameter. In other words, the
spacetime has a global
 event horizon.

Again there is the alternative of a naked singularity. In claiming that the
energy momentum tensor of the
 $Z$ is equal to the radiation in the $N$ minimal scalars, we have implicity
assumed that the radiation
 is uniquely determined by the conservation equations, the trace anomaly and
the boundary conditions at
 infinity. This will not be the case if there's a singularity visible from
${\cal I}_w^+$. So again the
 requirement that the radiation has positive energy implies there is either an
event horizon or a singularity.
 The arguments about loss of quantum coherence are then the same as in section
2.

\beginsection 4. Conclusions

 It is possible that two dimensional black holes are not a good model for the
four dimensional case. The fact that
 the field equations of the CGHS model with back reaction become singular at a
critical value of the dilaton field,
 suggests that this may be the case. However, if two dimensional models are any
guide to the real world, our results
 indicate that any Lorentzian description of black hole evaporation must have
horizons, or naked singularities,
or both. Of the two possibilities, naked singularities, would seem the worse.
Unless one has some boundary condition
 at a naked singularity, one can not predict what will happen. There is no
obvious candidate for such a boundary
 condition: the boundary conditions that have been proposed seem rather ad hoc.

 By contrast, in a Euclidean treatment, there is a natural boundary condition,
namely the no boundary condition,
 which says that there are no singularities and no boundaries in the Euclidean
domain, other than asymptotically
flat space. This boundary condition of no boundaries should mean that is
asymptotic Green functions are defined
 by a path integral over all fields and Euclidean metrics that are
asymptotically flat. These Green functions can
 then be used to calculate how ingoing particles evolve to outgoing particles,
maybe with loss of quantum coherence.
 It is not obvious that this process will have a Lorentzian description, but if
it does, our results suggest that
it will contain horizons.   By CPT symmetry, one might expect that there would
be past horizons as well as future
horizons. It is bad enough to lose quantum coherence, but to lose CPT symmetry
as well seems like
carelessness. This leads to a picture in which particles would fall into a hole
that was already existing in the vacuum.
 The hole would grow in size and mass and then evaporate down to a hole like
those in the vacuum. One might claim
 that the information about the particles that fell in was not lost, that it
was still contained in the residual
 black hole. But if this residual hole was indistinguishable from holes in the
vacuum, the information is
effectively lost, and the outgoing radiation will be in a mixed state.

 This picture is similar to that of scattering off an extreme magnetically
charged black hole: the hole grows
 in mass and then evaporates back to the original zero temperature black hole.
One can see that the information
 is contained in the residual black hole, but that is just words. The amount of
information that can be fed in
is infinite, and there is no way the information can be recovered. Moreover, as
the radiation is emitted in a
weak field region, there is no reason to distrust the semi classical
calculations that indicate that it is in
a mixed state. It is this effective loss of quantum coherence that is the
physically important result, rather than
 any semantics about whether the information can be thought of as being
contained in some remnant.

 The only difference between the picture being suggested here, and the
magnetically charged case, is that one
 would have to imagine that the ground state with zero mass and conserved
charge also contained objects with zero
temperature future and past horizons. But this is just what there is in the
Lorentzian section of a Euclidean
wormhole[12]. Consider the Euclidean metric $$ds^2=\left(1+{a ^2\over
x^2}\right)^2 dx^2$$ This corresponds to two
 asymptotically
 Euclidean regions connected by a wormhole or throat of size $a $. However, the
Lorentzian section obtain by
 $x^4\rightarrow ix^4$ looks rather different. Its Penrose diagram is shown in
figure 3. It has an outer null
infinity ${\cal I} _o$ like  flat Minkowski space but now the light cone of the
origin has also been sent to
infinity to become an inner null infinity ${\cal I} _i$. The two null
infinities intersect in two two spheres
$I ^+$ and $I ^-$. This is the four dimensional analogue of the Penrose diagram
for the linear dilaton solution,
 which also has two infinities. This supports the idea that there is a close
connection between wormholes and the
 formation and evaporation of black holes. Particles and information falling
into black holes pass into another
 universe, and particles from that universe enter ours in the form of black
hole radiation. Further developments
 of this idea will be published elsewhere.
\bigskip\bigskip
This work was supported in part by the U.S. Dept. of Energy under contract no.
DE-AC03-81-ER40050.
 Part of this work was done while SWH was a Sherman Fairchild Distinguished
Scholar at Caltech.

\vfill \eject
\noindent {\bf References}
\item {1.} Hawking, S.W., Particle Creation by Black Holes, Comm. Math. Phys.43
(1975), 199.
\item {2.} Hawking, S.W., Breakdown of Predictability in
Gravitational Collapse, Phys. Rev. D14 (1976), 2460.
\item {3.} Hawking, S.W., Black Holes and Thermodynamics, Phys. Rev. D13
(1976),191.

\item {4.}  Callan, C.G., Giddings, S.B., Harvey, J.A.,
Strominger, A. Evanescent Black Holes, Phys. Rev. D45 (1992), R1005.

\item {5.} Russo, J.G., Susskind, L., Thorlacius, L., Black Hole Evaporation
in 1+1 Dimensions, Phys.Lett. B292, (1992), 13.

\item {6.} Banks, T., Dabholkar, A., Douglas, M.R., O'Loughlin, M., Are
Horned Particles the Climax Of Hawking Evaporation?, Phys.Rev.D45 (1992) 3607.

\item {7.} Bilal,A.,Callan,C.G., Liouville Models of Black Hole
Evaporation, PUPT-1320, May 1992.

\item {8.} Russo, J.G., Susskind, L., Thorlacius, L. The
Endpoint of Hawking Evaporation, Phys.Rev.D (1992) 3444.

\item {9.} De Alwis, S.P., Quantum Black Holes in Two Dimensions,
Phys.Rev.D46 (1992) 5429.
\item {10.}  Hawking, S.W. and Stewart, J.M., Naked and
Thunderbolt Singularities in Black Hole Evaporation, to appear in Nucl.Phys.B.

\item {11.} Lowe, D., Semi-classical Approach to  Black Hole Evaporation,
Phys. Rev. D47 (1993), 2446.

\item {12.} Hawking, S.W., Wormholes in Spacetime, Phys. Rev. D37 (1987), 904.
 \end